\documentclass[preprint,12pt]{elsarticle}

\usepackage{lineno,hyperref}
\usepackage{multirow}

\usepackage{graphicx}
\usepackage{amssymb}
\usepackage{lineno}
\usepackage[binary-units=true,per-mode=symbol]{siunitx}
\usepackage{booktabs}
\usepackage{float}
\usepackage{subfig}
\usepackage{eurosym}

\usepackage{textcomp}
\usepackage{makecell}
\usepackage{graphicx}
\usepackage{xcolor}
\usepackage{amsmath}
\usepackage{siunitx}

\modulolinenumbers[5]

\journal{Future Generation Computer Systems}

\begin{document}

\begin{frontmatter}

\title{A Citizen Science Exploration of the X-ray Transient Sky using the EXTraS Science Gateway}

\author[cnr]{Daniele D'Agostino\corref{mycorrespondingauthor}}
\address[cnr]{National Research Council of Italy - CNR-IMATI, Genoa, Italy }
\cortext[mycorrespondingauthor]{Corresponding author}
\ead{dagostino@ge.imati.cnr.it}

\author[lei]{Duncan Law-Green}
\author[lei]{Mike Watson}
\address[lei]{Dept.of Physics \& Astronomy University of Leicester, U.K.}

\author[iuss,inaf]{\\Giovanni Novara}
\author[iuss,inaf,infn]{Andrea Tiengo}
\address[iuss]{Scuola Universitaria Superiore IUSS Pavia, Italy}
\address[inaf]{National Institute for Astrophysics – INAF, Milan, Italy}
\address[infn]{Istituto Nazionale di Fisica Nucleare - INFN, Italy}

\author[inaf]{Stefano Sandrelli}
\author[inaf]{\\Andrea Belfiore}
\author[inaf]{Ruben Salvaterra}
\author[inaf,infn]{Andrea De Luca}

\begin{abstract}
Modern soft X-ray observatories can yield unique insights into time domain astrophysics, and a huge amount of information is stored - and largely unexploited - in data archives.
Like a treasure-hunt, the EXTraS project harvested the hitherto unexplored temporal domain information buried in the serendipitous data collected by the European Photon Imaging Camera instrument onboard the XMM-Newton satellite in 20 years of observations.
The result is a vast catalog, describing the temporal behaviour of hundreds of thousands of X-ray sources. But the catalogue is just a starting point because it has to be, in its turn, further analysed. 
During the project an education activity has been defined and run in several workshops for high school students in Italy, Germany and UK. 
The final goal is to engage the students, and in perspective citizen scientists, to go through the whole validation process: they look into the data and try to discover new sources, or to characterize already known sources. This paper describes how the EXTraS science gateway is used to accomplish these tasks and highlights the first discovery, a flaring X-ray source in the globular cluster NGC 6540.

\end{abstract}


\end{frontmatter}


\section{Introduction}

X-ray astronomy probes a wide diversity of phenomena, related to the most extreme physical conditions that can be observed in the Universe: very strong gravitational and/or electromagnetic fields, very high temperatures, populations of particles moving close to the speed of light \cite{xray}. Almost all cosmic X-ray sources, from flaring stars in the solar neighbourhood, to accreting supermassive black holes in galactic nuclei at cosmological distances, are variable as a function of time and characterized by flux and spectral changes on distinct time scales, ranging from a fraction of a second to several years. Indeed, variability studies are crucial to understand the nature and physics of the sources.

Every day observing facilities with time-resolved imaging capabilities \cite{imagesec} collect huge amounts of potentially interesting information about serendipitous\footnote{lying by chance in the instrument field of view, in the portion of the sky surrounding the target of the observation} X-ray sources and their temporal variability, which remains mostly unused, stored in data archives. 
In the soft X-ray range ($\sim0.1-12$ keV), the European Photon Imaging Camera (EPIC)  instrument onboard the European Space Agency mission XMM-Newton is the most powerful tool to study the variability of faint X-ray sources, thanks to its unprecedented combination of large effective area, good angular, spectral and temporal resolution, and large field of view \cite{strueder,turner}. Twenty years after its launch, EPIC is still fully operational and its immensely rich archive of data, the XMM-Newton Science Archive\footnote{http://nxsa.esac.esa.int/nxsa-web} (XSA), keeps growing. 

Major efforts are ongoing to explore the serendipitous content in XMM data. For instance, a catalogue including all sources detected in EPIC observations is being regularly updated, as more data become available, and its most recent release\footnote{https://heasarc.gsfc.nasa.gov/W3Browse/xmm-newton/xmmssc.html} is the richest catalogue of X-ray sources ever produced. However, temporal domain information is not systematically extracted. The first, complete study of time variability in EPIC sources was carried out by the European-funded ``Exploring the X-ray Transient and variable Sky'' project (EXTraS, http://www.extras-fp7.eu) \cite{extras1}, which was run from 2014 to 2016 by several scientific institutions led by INAF, the Italian National Institute for Astrophysics: CNR-IMATI and IUSS Pavia (Italy) MPE and Friedrich-Alexander-Universit\"at Erlangen-N\"urnberg (Germany) and University of Leicester (UK). Its main goal was the realisation of a vast catalogue to be released to the astronomical community, describing and quantifying in a set of synthetic parameters the temporal behaviour of all EPIC sources. 
To accomplish this goal, new data analysis algorithms were designed and implemented. The results and software are available through a science gateway (http://portal.extras-fp7.eu) \cite{portal} made up of an open access database, named the EXTraS Archive,  and a portal for running the software on new data or to re-analyse an observation with different parameters. The software is of particular importance for enhancing the discovery potential of the XMM-Newton mission \cite{mus}. This is especially true as we look to the future: the EPIC instrument is still fully operational and collects new data daily, and its operations may be extended for more than one decade.

As the most sensitive search for variability ever performed, EXTraS is raising new questions in high-energy astrophysics \cite{science} and may serve as a pathfinder for future missions. The output catalogue, together with our analysis tools, discloses a huge amount of information, relevant for studies of almost all classes of astrophysical sources. The present effort is focused on delving into this vast dataset and finding interesting phenomena, as \cite{extras3,extras2}. 

Since the beginning of the project, project partners believed this would be an exciting challenge for citizens interested in astrophysics and student internships. Therefore an educational activity was defined for offering high school students the chance to go through the whole validation process: they study the data and try to discover new sources, or better to characterise already known sources \cite{extras5}. Apart from good basic science they can learn and use statistics, physics, image handling, the added value is that they could be the very first people in the world to discover and characterize those transient and variable sources: they managed to get their faces on the XMM-Newton cameras for a handful of seconds, just like extras in movies. Since October 2015 several workshops and related educational activities have taken place. This paper describes how  the EXTraS science gateway has been used together with the first important discovery, an enigmatic flaring source in the globular cluster NGC 6540  \cite{nature}.

The paper is organized as follows. Section \ref{sec2} discusses work related to citizen astronomy, Section \ref{sec3} presents the EXTraS portal while \ref{sec4} presents the EXtraS archive. Section \ref{sec5} describes the activities carried out for engaging students, while Section \ref{sec6} presents the conclusions.

\section{Citizen Astronomy and its Implementation}
\label{sec2}

As reported by the Encyclopaedia Britannica\footnote{http://www.britannica.com/EBchecked/topic/1993687/Citizen-Science-A-Platform-for-Nonprofessionals-Year-In-Review-2014}, ``The spectrum of projects and initiatives that fall under the umbrella of citizen science is broad, and consequently there is debate surrounding a formal definition''. 

One possible definition is ``A citizen scientist is a volunteer who collects and/or processes data as part of a scientific inquiry. Projects that involve citizen scientists are burgeoning, particularly in ecology and the environmental sciences, although the roots of citizen science go back to the very beginnings of modern science itself'' \cite{silverton}. 

The most famous - if not the first - initiative in this field is SETI@Home \cite{seti2}, which exploits the idle computing time of common citizens in the search for extraterrestrial life since 1999 \cite{nature1}. Presently, there are several  Internet-based citizen science projects and associations devoted to promote such collaboration paradigms, such as the Citizen Science Alliance\footnote{http://www.citizensciencealliance.org}, the OpenScientist blog\footnote{http://www.openscientist.org} and the  American Association for the Advancement of Science\footnote{http://www.aaas.org/}.
Furthermore publications using data and results collected with the contribution of citizen scientists are becoming common in many disciplines, as \cite{cspaper1, cspaper2, cspaper3}. 

Although a valuable motivation for a citizen scientist in joining a science project is the satisfaction of his/her intellectual curiosity, thus contributing to scientific dissemination \cite{know}, both the previous definitions as well as the cited activities and publications mainly stress the active role that citizens fulfill when participating in a project \cite{haklay}. They can:
\begin{itemize}
\item provide data, as for example in the Great Sunflower Project\footnote{http://www.greatsunflower.org/} or CrowdHydrology \cite{crownh}; 
\item provide computational resources to elaborate data, for example with the BOINC\footnote{http://boinc.berkeley.edu/} platform;
\item participate in the data processing and analysis.
\end{itemize}

The last two items represent the initiatives for involving citizen scientists in astronomy\footnote{https://www.iau.org/public/themes/citizen-science-projects/} \cite{rw1}. In particular there are several such initiatives, as can be seen in the Spacehack.org website, on the Zooniverse platform or - as regards providing computational resources - among the BOINC projects. 

Focusing on the data analysis, one of the most important platforms is represented by Zooniverse \cite{zoo1,zoo2}, ``a platform for people-powered research where anyone can study authentic objects gathered by researchers, like images of faraway galaxies, by answering simple questions about them''\footnote{https://www.zooniverse.org/about}. This platform hosts several projects, whose creation is straightforward, just requiring the upload of the data and a choice of the tasks the volunteers must do, such as answering questions or marking features in the data by drawing or tagging text\footnote{https://help.zooniverse.org/getting-started/}. The platform today hosts about 21 live projects related to astronomy \cite{zoo3,zoo7}, resulting in an increasing number of scientific discoveries, e.g. \cite{zoo4,zoo5,zoo6}. 

While Zooniverse is a collaboration between institutions from the U.K. and the U.S.  supporting the creation of science gateways with the specific object of offering citizen science tools, CitizenScience.gov is an official U.S. government initiative aiming at supporting federal agencies/institutions in accelerating the use of crowdsourcing and citizen science.  This means the focus is on providing a toolkit for improving  scientific projects software, i.e. to improve tools originally developed for professionals also to citizens, and on disseminating the advantages of this approach  by providing an extensive list\footnote{https://www.citizenscience.gov/catalog} with more than 430 successful active or past projects.

Classification and annotation of data represent the most common tasks, but more complex analyses can be performed. For example the  SkyServer initiative hosts several projects for supporting the study of stars and galaxies in the Sloan Digital Sky Survey (SDSS) \cite{skyserver}, i.e. the same objects that professional astronomers study. The focus here is twofold: to support the study and the use of heterogeneous data, e.g. images, spectra, photometric and spectroscopic data, and tools, e.g. for creating astronomical finding charts for an object seen by the SDSS or navigating the sky\footnote{http://skyserver.sdss.org/dr7/en/tools/started/}. Also Zooniverse is developing an environment where more advanced data analysis tools can be developed and provided to citizen scientists\footnote{https://blog.zooniverse.org/2013/11/26/zootools-going-deeper-with-zooniverse-project-data/ and https://github.com/zooniverse}.

A key factor for the success of such advanced initiatives, as outlined in \cite{rw1}, is represented by  ``providing user-friendly, web-based tools enabling fairly sophisticated data analysis to be performed by anyone with a browser. Open source tool code is a minimal requirement in this model; finding ways beyond this to support citizen algorithm development seems to be likely to pay off''. One of the most important examples in this direction, that can be represented as an evolution of SkyServer \cite{skyserver2}, is the SciServer collaborative environment for large-scale data-driven science \cite{sciserver}. It uses Jupyter notebooks \cite{jp} for  advanced analysis in a Cloud environment. Jupyter is an interactive computing environment based on the concept of notebooks, i.e. documents including text, plots, equations, videos but also live code and interactive widgets written in Python, R or other languages. In particular the output generated by running code is embedded in the notebook, which makes it a complete and self-contained record of a computation that can be executed locally or on the same server as the source data. The result is an environment able to support highly complex searches and analyses resulting in  millions of sky objects from several databases\footnote{http://www.sciserver.org/integration/astronomy/}.

Moreover Jupyter notebooks can be used also for creating science gateways \cite{sg2}. In general, a science gateway  can be defined as a set of software, data collections, instrumentation and computational capabilities that are integrated - using different technologies and middleware \cite{airavata} - via a Web portal (or a desktop application) in a user friendly and effective environment supporting the scientific research and education activities of a specific community. Science gateways are gaining an increasing interest in many communities \cite{sg, drihm}, such as the astrophysics one \cite{visivo, becciani2, astro3}, also because one of the best strategies 
to provide software and data to a scientific community is through a set of services designed following this paradigm. The portal we developed in the project and used for involving citizen scientists \cite{portal} is based on this paradigm.

In conclusion most of the presented citizen science projects represent science gateways accordingly to the definition we reported. However, these gateways normally offer to citizens simplified interfaces with respect to those offered to scientists, in order to balance the efforts in their learning curve with the need to provide them enough flexibility to actually carry out effective analyses\footnote{https://daily.zooniverse.org/2017/09/11/keeping-it-simple/}. But, at the same time, projects like  SciServer allow college students and and science-literate citizens to access the same tools as researchers who work with Big Data\footnote{http://www.sciserver.org/about/vision/}.

\section{The EXTraS Portal}
\label{sec3}

\begin{figure*}[!hbt]
\centering
\includegraphics[width=\textwidth]{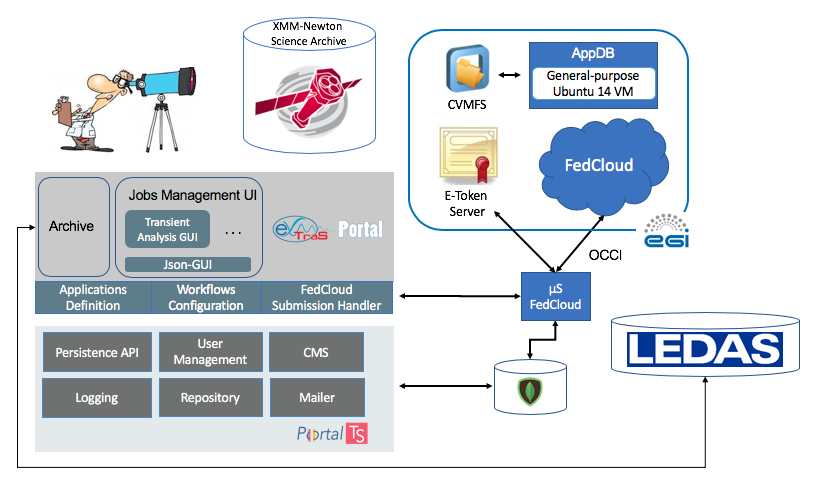}
\caption{The architecture of the EXTraS science gateway}
\label{fig_arch}
\end{figure*}

The EXTraS science gateway is composed of an archive, presented in the following Section, and a portal. The architecture is depicted in Figure \ref{fig_arch} and described in details here and in the following Section.

The EXTraS portal has the goal of providing users with a seamless environment to process the observations made available from the XSA with the EXTraS pipelines, hereafter called experiments. It provides an user-friendly interface, with no need for the installation of any software. 

\paragraph*{Experiment configuration}
The Jobs Management User interface represents the home page of the EXTraS portal. It provides users with the possibility of creating, submitting and managing the different analysis experiments based on the software developed within the EXTraS project. In particular it presents all the submitted or configured analyses, providing the possibility to create a new analysis starting from an existing configuration or share results with other users.

This module is based on AngularJS and it is a complete web app, without any server side code \cite{gui}. It uses the Persistence API to store and retrieve experimental data, and it activates the other portal modules corresponding to the different operations available.

\begin{figure*}[!hbt]
\centering
\includegraphics[width=\textwidth]{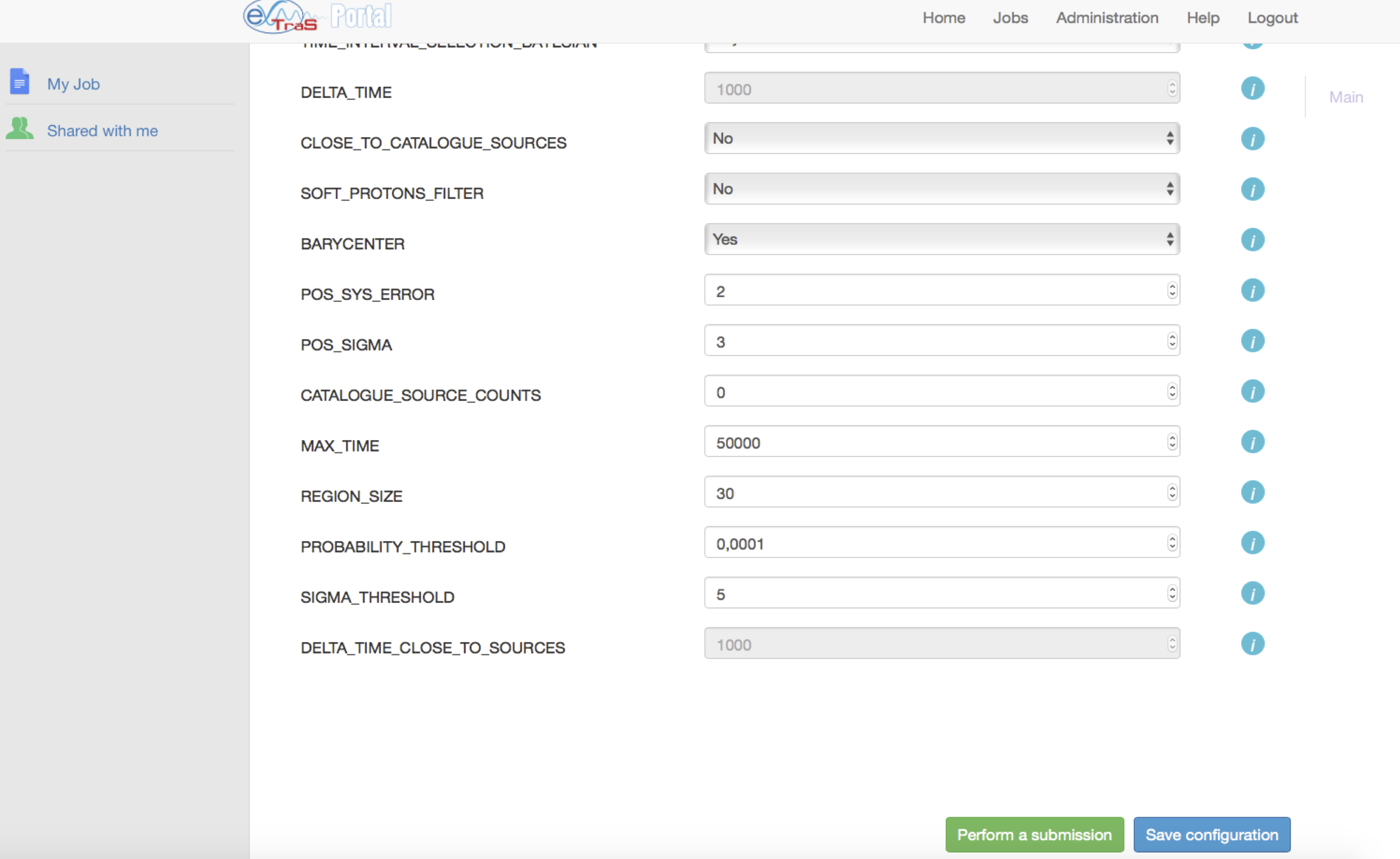}
\caption{The Transient Analysis UI shown by the Workflow configuration module of the EXTraS portal.}
\label{fig_wc}
\end{figure*}

\paragraph*{Experiment execution}
All jobs are managed by the portal, based on computing
resources provided by EGI Fedcloud \cite{fedcloud} to virtual organizations (VO), i.e.groups of users where members are usually in related research activities. The Workflow Configuration module is responsible for interacting with the user for the creation and configuration of experiments based on the EXTraS software. In the portal every analysis corresponds to a single application, therefore there is no need to explicitly create and manage workflows. For this reason a user interacts directly with an analysis-specific  interface (UI), as shown in Figure \ref{fig_wc} for the Transient Analysis. Its main aim is to collect the parameters value and to create a namelist, that will be provided to the FedCloud Submission Handler for the actual execution of the job. 

The FedCloud Submission Handler module manages submission of jobs  and provides a full view of the job status, results and logs. In particular the actual submission requires the user specifies as input for the analysis one or more observation identifiers (OBSID) among those contained in the XSA. Each of them corresponds to a job, therefore an analysis configuration can result in multiple jobs executed at a time.  During the execution the user can monitor the status of the job by means of the real-time log information the software tool provides.

Since its official release at the beginning of 2017, 54 users registered to the portal and submitted experiments for a total of 32,240 core-hour in 2017, 505,380 in 2018 and 10,040 in 2019, as reported by the EGI Accounting portal\footnote{ https://accounting.egi.eu/vo\_admin/cloud/}.

\paragraph*{Results management}
When a job terminates, the FedCloud Submission Handler  provides the possibility to retrieve results and also log information. 
All the information related to a job (e.g. the configuration parameters, the logs, the results, the ownership/sharing information and possible comments) are stored on the portal database via the Persistence API until it is deleted by the user who owns it. 

The  EXTraS portal provides two further key features: the ability to share an analysis (i.e. the namelist and possibly the results) and support for the interaction and discussion (in terms of comments) among the users sharing it.
Sharing a completed job means not only that the experiment results are visible to other users, but also that the configuration is shared and can be used as a starting point for re-submitting the experiment on a new set of data. Thus, a job execution can be replicated by other users that can, for example, validate the experiment results or explore the behaviour by changing one or a few parameter values. 

Any result computed within the portal is not automatically transferred in the archive, but it has to be validated by the project community, that can use the portal to publicly discuss it.

\paragraph*{Non-academic users}
Citizen scientists can interact with the portal exactly as the scientists can. In particular they can exploit the EGI Fedcloud infrastructure for submitting jobs. This has been made possible by the use of ``robot certificates'' \cite{robot}.

EGI Fedcloud in fact relies on a single sign-on mechanism to access the federated services based on X.509 certificates and VO membership. Robot certificates were introduced to allow users, who cannot get or are not familiar with personal digital certificates, to exploit any distributed infrastructure relying on them in their research activities. The robot certificate is usually associated with a specific application (or function) that the application developer/provider wants to share with all the VO \cite{robot2}. This is exactly the scenario which has arisen in the EXTraS activities, because portal users are provided with the  possibility to run only pre-defined software tools. Therefore the FedCloud Submission Handler module interacts with an e-Token server \cite{etoken} to submit jobs on the resources granted to the ``extras-fp7.eu'' VO \cite{egisla}.

Furthermore, the support for the interaction and discussion  is very important also for the activity of citizen scientists, that can be assigned a set of XSA observations and parameters to inspect. In this case they can share the results and a preliminary analysis among them and with the scientists, for a fast but effective screening.

\section{The EXTraS Data Archive}
\label{sec4}
The EXTraS Data Archive forms the primary online repository for the data generated during the
EXTraS project. This builds on the pointed and slew XMM-Newton Serendipitous Source Catalogues \cite{saxton, rosen} and supports a wide range of newly-generated X-ray light curves, associated source catalogues with measures of variability and statistical significance, and derived products such as hardness ratios, power spectra, source classifications and source summary files.

The design of the archive is derived from the existing high-energy astrophysics data archive Leicester Database and Archive Service (LEDAS\footnote{ https://www.ledas.ac.uk}) at the University of Leicester, which hosts data from several major X-ray missions, including a full set of XMM-Newton EPIC serendipitous data (around 200 million seconds of observing time). The core archive system originally developed for LEDAS has been completely rewritten to current software development standards (including Model-View-Control architecture and unit testing). 
Moreover a major technological development in archive provision post-2000 has been the rise of e-Science, and particularly in the field of astronomy, the Virtual Observatory (VObs) concept \cite{vo1, vo2}. The goal of the VObs is to provide a standardised set of metadata, access protocols and supporting software tools to improve access to, and interoperability between, very large astronomical datasets.

\paragraph*{The data}
EXTraS draws on the 3XMM-DR4 public data to search for periodic, aperiodic and transient sources with much greater fidelity than the basic light curves and Fourier Transforms generated as part of the 3XMM automated processing. This variability is being studied down to the native time resolution of the instruments (typically 73 ms and 2.6 s for the EPIC pn and MOS cameras respectively). Long-term variability is being characterised for all sources with multi-epoch data. Upper limits will be computed in the case of non-detections in specific observations.

Catalogues and bulk products produced during the EXTraS project have been
transferred to the University of Leicester, UK (UoL) to be incorporated in LEDAS. Details are provided in \cite{duncan}. A total of 18 TB of LEDAS/EXTraS data is presently held on UoL central archival storage.

In order for VObs applications to handle EXTraS time series datasets fully and
consistently, the data structures and metadata must be documented according to a standard
syntax, as defined by the International Virtual Observatory Alliance (IVOA) \cite{Graham}.
But the existing data models are insufficient to describe EXTraS data products.
This is the reason why during the project we consider that the development of a specific data model  should follow two basic principles: simplicity, which eases management and adoption, and adherence to existing IVOA standards, which increases the probability of wider adoption. To that end, we propose to base the new data model, named ExtDM, on the existing PhotDM standard, with extension to the time domain drawn from the lower-level axis definition of the Characterisation DM (CharDM) \cite{duncan}.

We joined the standards group for time series data within the IVOA with the aim of advancing the interests of the EXTraS project, and incorporating any approved standards into the EXTraS infrastructure at the earliest opportunity. The process is ongoing at the time of writing.

\paragraph*{The search form}

The main page of the archive provides a top-level menu to access EXTraS data products subdivided by the kind of analysis they are related to, i.e. Aperiodic Variability, Search for Periodicity, Transient \& Highly Variable Sources, Long-Term Variability and Multiwavelength classification. 
A combined catalogue, which searches all EXTraS source catalogues, and a basic catalogue cross-matching facility are also provided. Direct catalogue downloads in FITS \cite{fits} and VOTable \cite{votable} formats are available. It also provides an option to access all catalogues currently held by the LEDAS system (over 900 catalogues at the time of writing) covering a wide variety of missions and wavebands. Online help is available for all catalogues.

\begin{figure*}[!hbt]
\centering
\includegraphics[width=\textwidth]{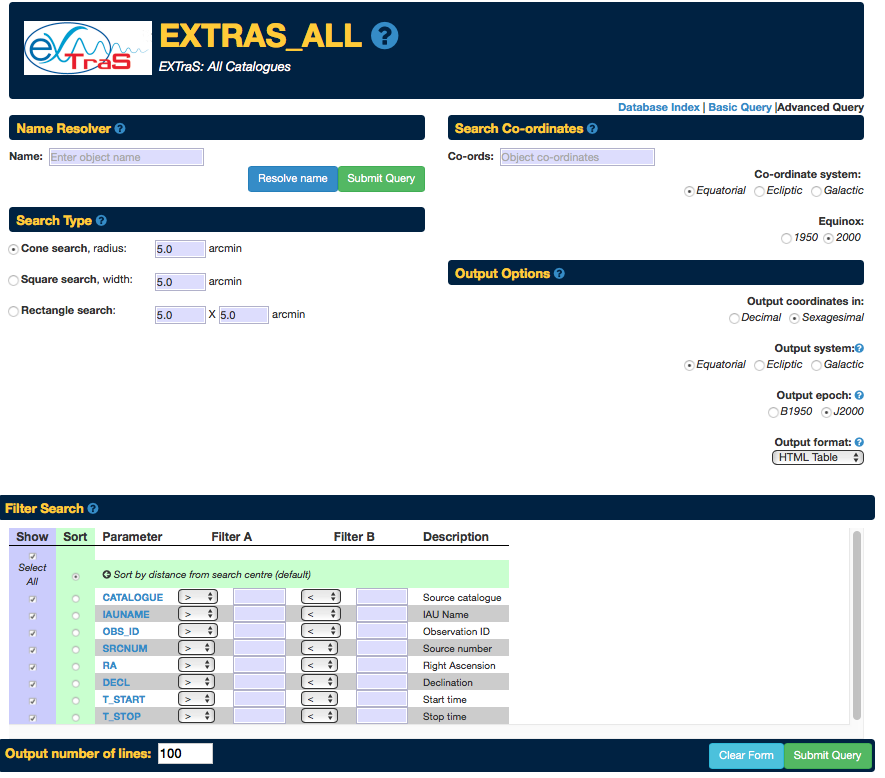}
\caption{The advanced catalogue search form in the EXTraS archive}
\label{fig_db1}
\end{figure*}

The catalogue search form, shown in Figure \ref{fig_db1}, allows searches on a given sky position, using either a cone (fixed radius from search centre), box or rectangle search area.
The form makes also possible to type the source name and click on ``Resolve Name''. In this case the Simbad database \cite{sinbad} is queried to find the source coordinates for the object. If successful, these are displayed in the Cone Search input box on the query form.
A minimal or full set of table columns can be selected, and a variety of output formats displayed.

\begin{figure*}[!hbt]
\centering
\includegraphics[width=\textwidth]{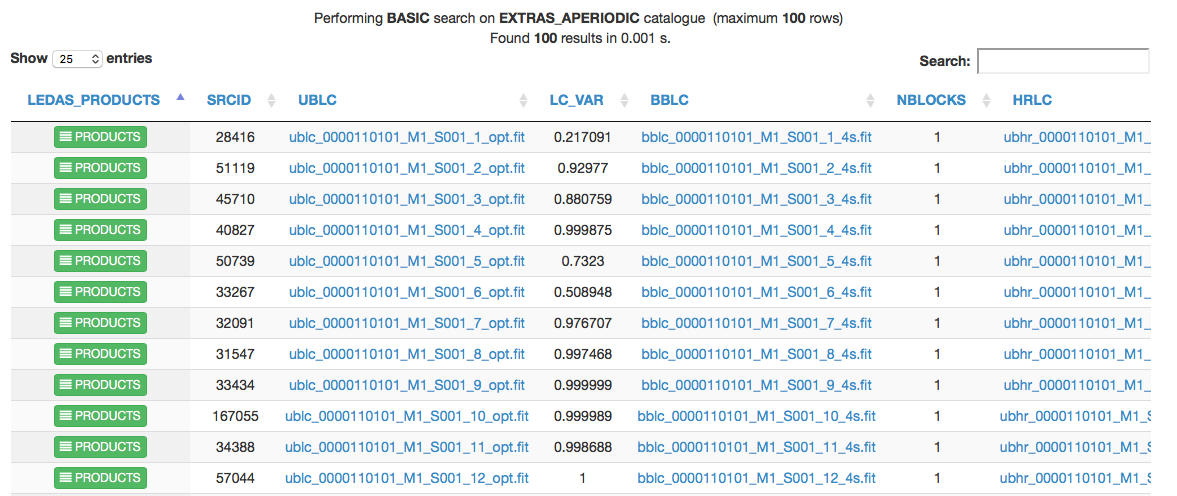}
\caption{The full list of results as an HTML page. The buttons on the left allow one to explore all the products and the links to directly download selected files}
\label{fig_db2}
\end{figure*}

\begin{figure*}[!hbt]
\centering
\includegraphics[width=\textwidth]{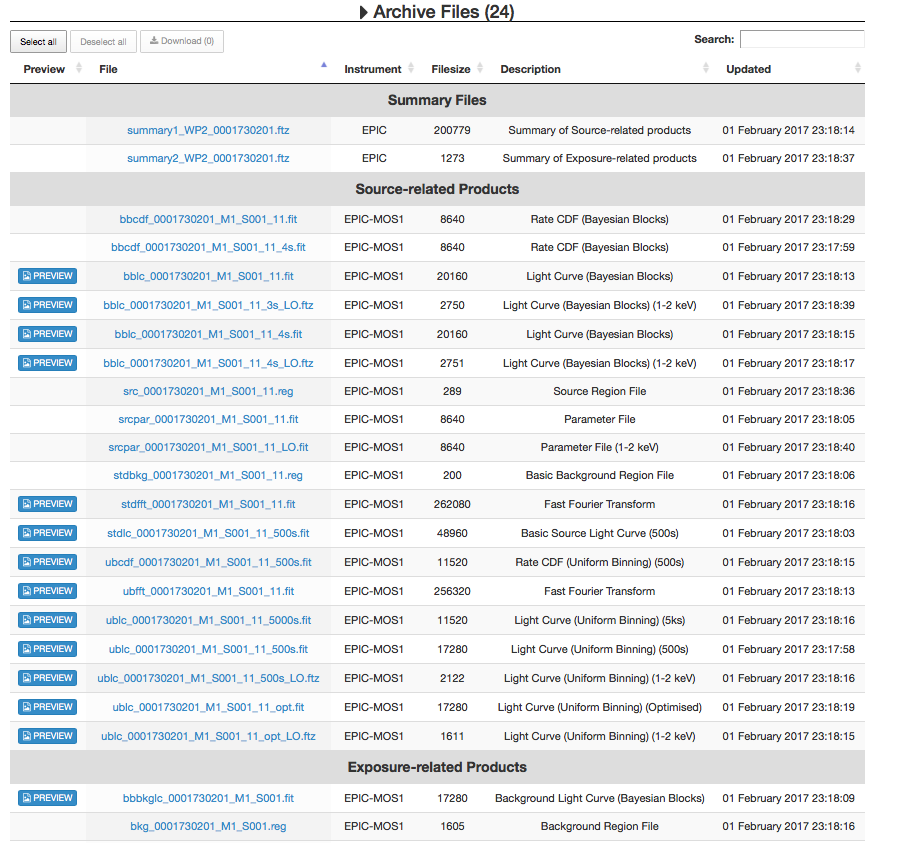}
\caption{The summary page of the catalogue results for each single source}
\label{fig_db3}
\end{figure*}

Generation of archive search results in multiple formats (HTML, ASCII table, CSV, VOTable
etc) is accomplished using the Twig templating language for PHP\footnote{http://twig.sensiolabs.org/}. Twig is a full-featured presentational language for data output including generation of web pages -- it handles branch and loop constructs and supports multiple inheritance, enabling complex page layouts to be built from a common set of simple blocks. Tabular search results are presented using the Javascript DataTables library\footnote{http://www.datatables.net/}, which produces tables which can be filtered, sorted and paginated on the fly by the end user. An example is shown in Figure \ref{fig_db2} for the full list of results, while the bulk products associated within each given dataset ID organised by category is shown in Figure \ref{fig_db3}.

\paragraph*{Visualization}
Beside the download of the results data, an expanding set of interactive visualisations for EXTraS catalogue and bulk product data is provided. The visualisations are generated directly in the browser (i.e. by clicking on the blue ``PREVIEW'' button at left links in Figure \ref{fig_db3})  and require no additional software installation.

The time-series analysis tasks generate science output typically in FITS format, which cannot natively be displayed in a web browser. A static ``quick look'' graphical view for e.g. spectra and time series is already available for sources present in the 3XMM-DR5 catalogue in PNG and PDF formats. For new sources the archive data ingest script is able to  generate such static images automatically using e.g. Python matplotlib.

\begin{figure*}[!hbt]
\centering
\includegraphics[width=\textwidth]{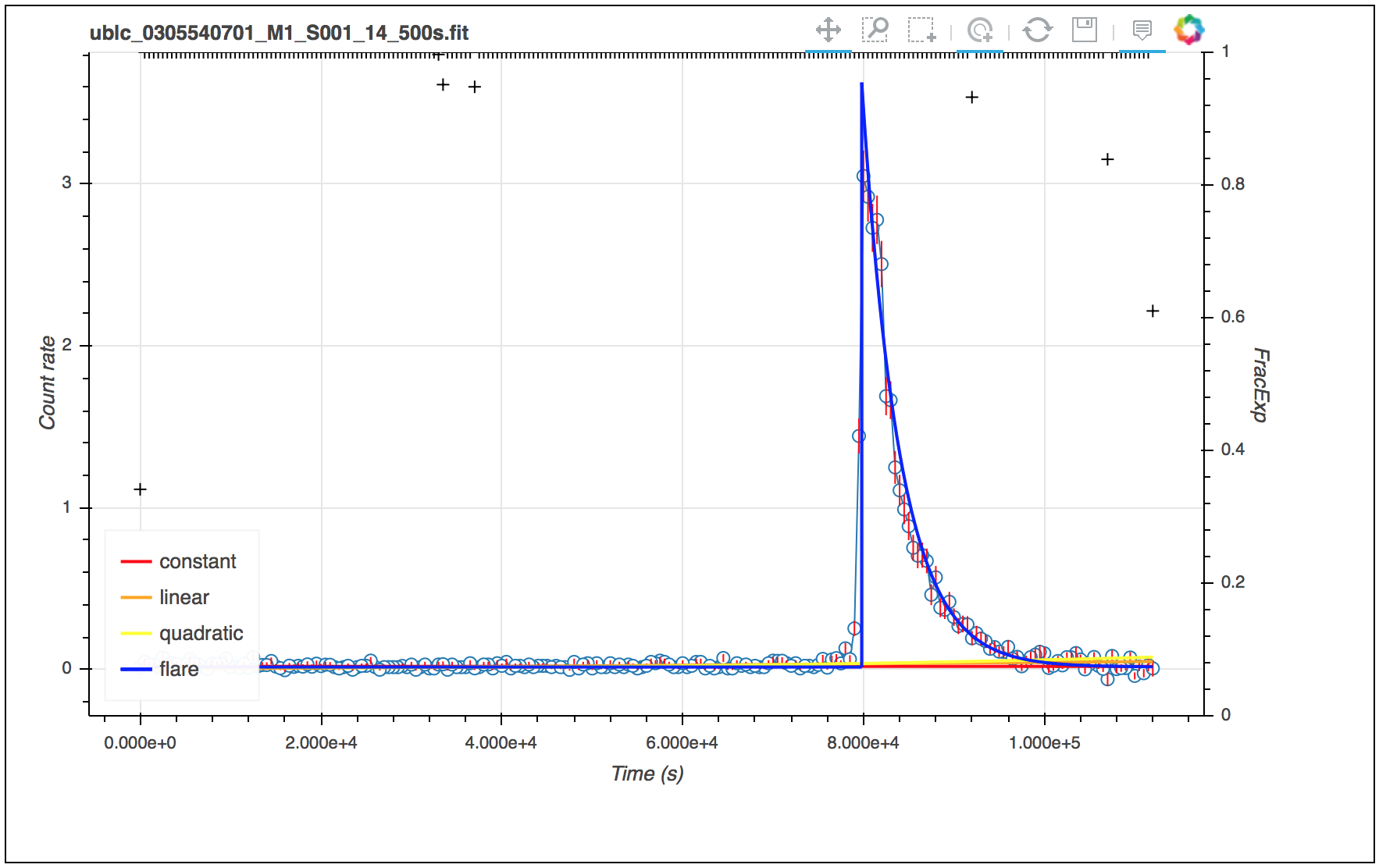}
\caption{Bokeh visualisation of Aperiodic Variability MOS1 timeseries}
\label{fig_vis1}
\end{figure*}

\begin{figure*}[!hbt]
\centering
\includegraphics[width=\textwidth]{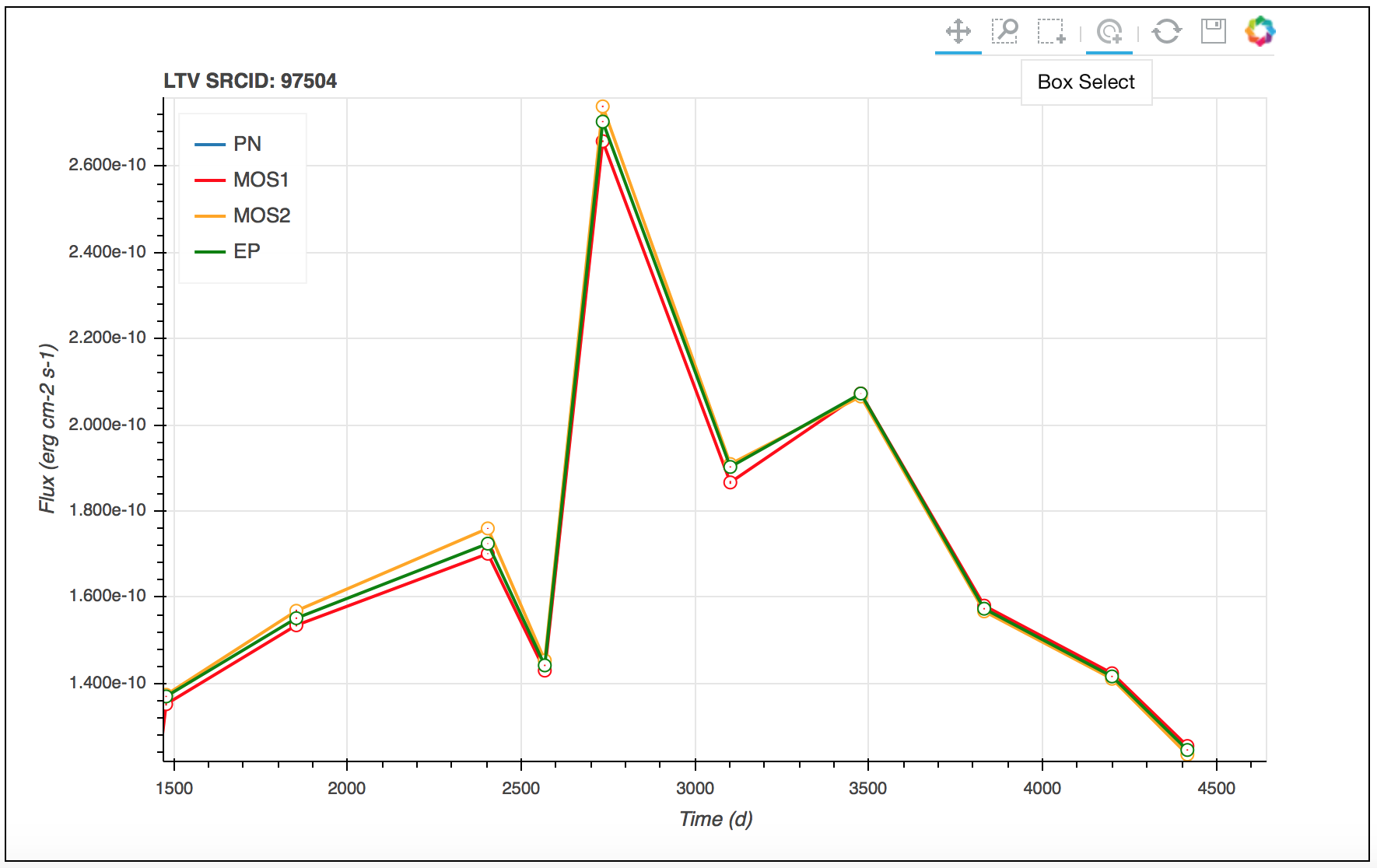}
\caption{Bokeh visualisation of Long-Term Variability timeseries}
\label{fig_vis2}
\end{figure*}

Moreover the archive interface has been enriched with dynamic, interactive visualisations using the Python graphics library Bokeh\footnote{http://bokeh.pydata.org/} running on the Python WSGI application server gunicorn\footnote{http://gunicorn.org/}. Requests are sent to the visualisation server using a REST interface, and the server responds with a JSON data block which is rendered by the view using standard AJAX techniques. Examples of EXT-PDA visualisation output are shown in Figures \ref{fig_vis1} and \ref{fig_vis2}.

The use of JSON for data transfer enables additional features to be added to EXTraS archive data analysis in the future using standard Javascript libraries, such as jStat\footnote{http://www.jstat.org}.

\section{The Education Activities}
\label{sec5}
During the project a workshop scheme has been designed to let high school students play as researchers, i.e. they are introduced to the scientific method and they learn how to work with the EXTraS software tools and results.
The formal goal of the students is the validation and classification of astronomical sources with a peculiar temporal behaviour (pre-selected by researchers), by exploring the data extracted from the EXTraS data archive. 

The workshop has been designed as a live role playing-like activity with a strong hands-on strategy and some formal education: the students are divided into cooperating groups that attend laboratories and lessons given by researchers.  
The groups gather information about high energy astrophysics, space technologies, data mining and analysis. Their ability to interact with experts is fundamental to reach their goals.

A typical workshop lasts about 1-2 weeks (40-80 hours).
We explicitly adopt an inquiry-based learning strategy, together with a peer-to-peer education approach. This global strategy both fosters the critical thinking and engages the students, who act as though they were researchers. 
A short but clear framework for describing inquiry in astronomy is provided by the IAU project astroEDU\footnote{http://astroedu.iau.org/ebl/} \cite{iaupub}.

During the analysis of possibly interesting X-ray source candidates they are guided by five major questions:
\begin{enumerate}
\item Is the selected candidate a real, astrophysical X-ray source?
\item Is the peculiar temporal behaviour related to any instrumental effect?
\item Has the source a possible counterpart at other wavelengths (e.g. optical)?
\item What kind of astrophysical object or phenomenon might it be?
\item Might it be an important scientific discovery?
\end{enumerate}

 We run several workshops and other education activities laying on the workshop experience in Italy, Germany and United Kingdom, engaging more than 200 high school students, from 2015 onwards. Details are available on the project website\footnote{http://www.extras-fp7.eu/index.php/educational/list-of-workshops}. 

The first part of the workshop aims at engaging participants in contemporary astrophysics, with a specific regard to the high energy band. Researchers or teachers introduce them to the contents and the technical language of astronomy (such as field of view, electromagnetic spectrum, time resolution, spatial resolution and energy resolution, light curve, photons and so on) and maths (such as probability, confidence level and so on). At this stage, researchers and teachers should try to avoid a frontal approach. They are requested to solicit questions and discussion on the treated topics, instead of acting as pure experts. The goal of this part is to provide the students with the right method and conceptual tools to face the tasks of data reduction and interpretation, so that some kind of passive learning is assumed on the part of the students.

The second part of the workshop aims at giving the students the opportunity to analyse the results produced by the EXTraS software tools, either stored in the archive, or generated within the workshop by using the portal. Different analysis tasks have been implemented, based on the use of different resources and tools, and related to a large diversity of astrophysical phenomena.

\paragraph*{Using the archive}
A sample of sources with peculiar temporal behaviour is pre-selected by researchers, based on a blind search in the results from the automatic characterization of variability performed within EXTraS. Such sources are proposed to the students for validation and classification. We considered the case of short-term variability, ranging from a few seconds to a few hours (i.e. on a time scale shorter than the duration of an XMM observation) -- in particular, focusing on flaring/transient sources (displaying a sudden, large increase in flux, and a subsequent fading to the pre-burst flux level) and on eclipsing/dipping sources (displaying a sudden drop in flux, followed by a fast rise to the pre-eclipse level). For each source, students are asked to:

\begin{enumerate}
\item confirm that the source candidate corresponds to a real astronomical source by inspecting the EPIC sky images (is the source candidate close to a CCD edge or defect? Is it located close to a very bright source or superimposed to a bright diffuse source?);
\item confirm that the candidate flare or dip is real, by inspecting the light curves of the three EPIC cameras;
\item exclude the possibility that the flare/dip is related to imperfect background subtraction, by inspecting the background light curves (is the feature simultaneous to a background flare?);
\item check whether the feature can be detected in multiple energy ranges, by inspecting energy-resolved light curves;
\item check whether the source is already known and classified, and whether its peculiar temporal behaviour had already been observed and published, by using the Simbad database;
\item if the source is not known/classified, search for possible counterparts in multi-wavelength catalogues and using the ESASky tool\footnote{http://sky.esa.int};
\item propose a classification for the source and an interpretation for its temporal behaviour;
\item try to determine whether this could be an interesting discovery.
\end{enumerate}

All of the above steps can be completed by inspecting EXTraS products in the archive, or using online resources linked to the archive web pages (e.g. the 3XMM catalogue and its products, the Simbad database, results of cross-correlation with multi-wavelength catalogues).

We decided to focus on the cases of flares and eclipses, because the simple shape of the temporal modulation eases several steps of the analysis. In principle, the same approach could be applied to any kind of temporal variability. For instance, an interesting possibility could be to study sources with large variability on a long time scale (ranging from few days to few years - e.g. between different XMM observations). Details on how the students use the archive interface are provided in \ref{append}.

\paragraph*{Using the portal}

The EXTraS science gateway has also been designed to give non-expert users the possibility of extending the EXTraS analysis to new XMM observations or reproducing the EXTraS analysis of old observations with different settings and input parameters. During the workshop, the students have also the opportunity to experience this additional step in the work of a researcher by selecting targets from recent XMM observations, not included in the EXTraS archive, and performing the full EXTraS analysis, starting from raw data and reaching down to the interpretation of results. 

The choice of the targets of the new observations is left to the groups of students, with only some general suggestions by the researchers. This resulted in some groups selecting particular sky regions (e.g., star clusters or nearby galaxies) and other choosing random positions (e.g., selecting the most recent or the longest observations).

After the selection of the XMM observations (the observation ID is the main input data that students enter into the portal interface),  the students could process the data (automatically downloaded from XSA) with the same Transient \& High Variability software developed and used in the project, with the possibility to modify a limited set of parameters (e.g., the detection threshold), using  the portal UI as described in details in \cite{portal}. In order to obtain the detection of a significant number of transient candidates to validate, the researchers suggested suitable values, but encouraged the students to repeat the analysis with different parameters.

Since the full data processing of each observation takes from a few to several hours, each group could select only a handful of observations and the data processing could be performed by submitting the jobs to Fedcloud resources and collecting the results on the following day.
After the data processing the results are stored in a TAR (Unix Tape Archive format) file ready to be downloaded to the students' laptops,  independently of the archive interface.
The results can then be analyzed by the students following a procedure similar to the one described before for the analysis of sources in the EXTraS archive. 

Given the limited number of observations that could be processed during the workshops, compared to the much larger amount of computing time used during the EXTraS project to populate the archive, only a few possibly real X-ray transients could be discovered by the students. 

This aspect is of particular importance. On the one side we showed the students a significant part of the scientific process leading to scientific discovery and the difficulty in obtaining scientifically relevant results. On the other side this limited test demonstrates how the use of the portal by a much broader community of citizen scientists, guided by expert astronomers, could substantially support the scientific community in performing such analysis.

Despite the low probability of finding a new transient, analyzing the observation 0781690201 a group of students discovered a new X-ray transient candidate.
This is an observation of a region of the Galactic plane with a duration of $\sim78$ ks.
The Galactic plane is characterized by the presence of numerous
stars that can give rise to stellar flares, which are
the most frequent transient events expected in the X-ray sky.
Using the visualization software developed within the EXTraS project for the screening analysis\footnote{http://www.extras-fp7.eu/index.php/scientific-community/softwaretools}, the students identified the new X-ray transient by analyzing the light curve of the three EPIC instruments, selecting the 0.5-4.5 keV energy band.

\begin{figure*}[!hbt]
\centering
\includegraphics[width=\textwidth]{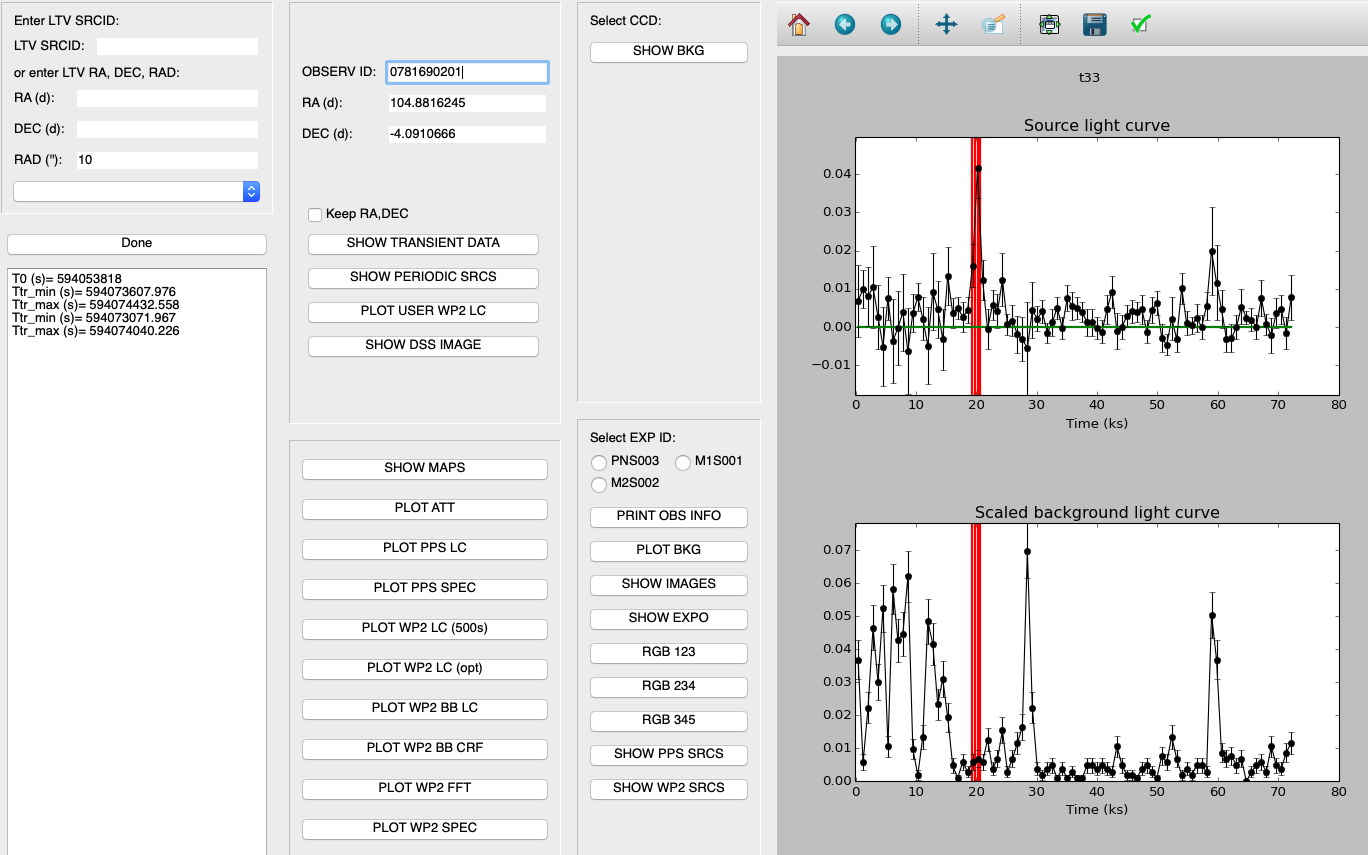}
\caption{The analysis of a transient candidate with the EXTraS screening visualization tool.}
\label{fig_sf}
\end{figure*}

The peak in the light curve is clearly visible in Fig \ref{fig_sf}, with a duration of $\sim800$ seconds, in a period of low background.
The students then entered the coordinates of the transient candidate in the ESASky\footnote{http://sky.esa.int} online tool to search for counterparts in other energy bands, finding a correspondence with a clearly visible star in the Optical/IR bands.

\paragraph*{The first result: the peculiar transient J1806-27}

A very unusual phenomenon was discovered within a workshop held at INAF/IASF Milano in September 2017. Six high-school students participated in the research work: Lorenzo Apollonio, Bartolomeo Bottazzi-Baldi, Martino Giobbio, Razvan Patrolea, Elena Pecchini and Cinzia Torrente from Liceo Scientifico G. B. Grassi, Saronno, Italy.

As an input for the students' work, we blindly selected $\sim200$ sources in the EXTraS database -- half of them being consistent with a transient behaviour, and half with dipping behaviour. The students were arranged in two groups, with the task of scanning the list of sources, following the steps described in the previous section.
At the end of the workshop, the two groups presented their results -- a small sample of sources they had selected as potentially interesting. The case of the candidate transient 3XMM J180608.9-274553 (J1806-27) drew our attention. 

\begin{figure*}[!hbt]
\centering
\includegraphics[width=\textwidth]{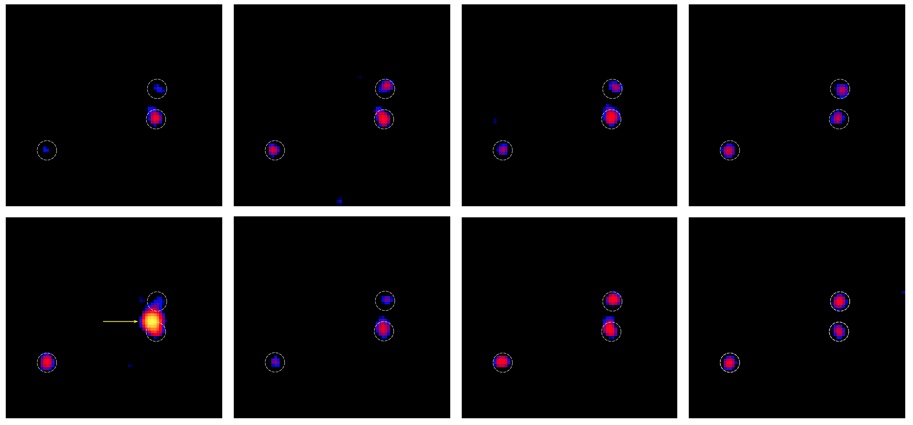}
\caption{The brightness of J180608.9–274553 changed by up to 50 times its normal level in 2005, and quickly fell again after about five minutes. }
\label{flare}
\end{figure*}

The transient is detected with high significance by all EPIC cameras and lasts a few minutes only, as shown in Fig. \ref{flare}. The source, which had never been studied before, lies within the core of the Galactic globular cluster NGC 6540. The students' report triggered a more detailed analysis by a team of astronomers at INAF/IASF. The transient, with a duration of $\sim300$ s and a symmetric rise/decay time, turned out to be very peculiar, defying any classification. Assuming the source to be in NGC 6540, its peak luminosity of $\sim10^{34}$ erg s$^{-1}$ is orders of magnitude lower than the one of thermonuclear X-ray bursts from neutron stars in low-mass X-ray binary systems; a flare from a coronally active star in the cluster can also be excluded because of the short duration, a factor $\sim100$ smaller than expected based on the known correlation between luminosity and duration of stellar flares. The possibility of a less luminous flare from a foreground star is also unlikely, because of the lack of an optical counterpart in archival images collected with the Hubble Space Telescope. The properties of this peculiar transient have been published in full detail in \cite{globular}. The discovery and the active role played by students were the subject of an article on the web site of the European Space Agency\footnote{https://goo.gl/rUigBv}, and were also included in the weekly ``Research highlights''  section by the Nature magazine\footnote{https://www.nature.com/articles/d41586-018-05959-4}.

\section{Conclusions and Future Work}
\label{sec6}

The EXTraS project, being related to rare, exotic phenomena and to extreme, poorly known astrophysical sources, has  great potential for the popularisation of science in general and of astronomy in particular. 

The researchers involved in the project deem outreach and involvement with schools and other institutions a fundamental component of their scientific mandate. Astronomy has a fundamental role in enhancing interest in science in young people, laying the basis for a career in Science, Technology, Engineering and Mathematics (STEM).
The outreach and involvement of people of all ages is very important as well, because it generates interest in scientific research and fosters the diffusion of scientific culture. 

Within this framework, the availability of a vast catalogue and of the appropriate software – both of them online and freely usable - allows us to think of the EXTraS activity as a Citizen Science candidate project, with a particularly interesting activity for the classroom.

This paper presented the EXtraS science gateway and the didactic activity schema for high school that could foresee: the extraction of transient candidates through the algorithm implemented by the EXTraS team; the validation procedure; the search for a counterpart; a final discussion of the results.

The success of the workshops held in these years shows that a citizen science approach can be very useful at school, even if the proposed activity is far outside the students'
curricula, provided that we offer them a stimulating challenge and framework. In addition, we showed that students can be engaged in a high level challenge. 
Indeed, the use of a Science Gateway as a citizen science platform is part of the challenge itself. At first sight, it can seem to be a complex tool to be used by students and teachers, such as an actual scientific laboratory. 
However, as soon as they become familiar with it, it rapidly turns into a chance to explore a whole scientific world: they can create, submit, manage different analysis experiments with their own choice of configuration parameters; they can retrieve and discuss the results and, moreover, they share parameters, data and results with other users (scientists or students), as we describe. This fosters the formation of a sort of enlarged and motivated user community, with a very strong effect of engagement of the non academic world – which is one of the current assets of science communication and life-long learning.

Our next step is to launch the activity as a totally citizen science one by improving existing didactic material for an on-line e-learning platform: we believe every interested citizen could be able to exploit the science gateway to identify and cooperate with us to analyze possibly interesting X-ray source candidates.

\section*{Acknowledgments}
This research has made use of data produced by the EXTraS project, funded by the European Union's Seventh Framework Programme under grant agreement no 607452. 
This work used the EGI infrastructure with the dedicated support of CYFRONET-CLOUD and INFN-CATANIA-STACK. Authors would like to thanks the EGI staff, in particular Giuseppe La Rocca and Diego Scardaci, for their valuable support.
J1806–27 was selected as a potentially interesting source by L. Apollonio, B. Bottazzi-Baldi, M. Giobbio, R. F. Patrolea, E. Pecchini and C. A. Torrente (Liceo Scientifico G. B. Grassi, Saronno) during their stage at INAF-IASF Milano in 2017, September, within the Alternanza Scuola-Lavoro initiative of the Italian Ministry of Education, University and Research.

\begin{appendix} 
\section{Study of candidate flaring sources}
\label{append}
Each group is assigned a list of pre-selected candidate flaring sources -- an ASCII file with one row per source, including the source sky coordinates and a small set of other  source identifiers (e.g. observation ID, exposure ID, unique source ID). For each source in the list, each group should complete the following steps, guided by the 5 major questions listed in Sect.~\ref{sec5}.

\begin{enumerate}

\item Go to the EXTraS Archive main data access page, select the ``Aperiodic, short-term variability'' query form and search the source by coordinates.

\item in the Results page, a list of all the XMM exposures (for each EPIC camera) analyzed by EXTraS is displayed. Select the appropriate observation and exposure ID for the pn camera and click on the Products button. 

\item a list of all bulk products generated by EXTraS for the pn camera in the selected exposure is displayed. Follow the link to the 3XMM summary page. Inspect images: is the source in a noisy/confused region? Is it close to a detector edge? 

\end{enumerate}

Step 3 allows one to answer the major question: ``Is the selected candidate a real, astrophysical X-ray source?'' 

\begin{enumerate}
\setcounter{enumi}{4}

\item Go back to the list of bulk products; select the Bayesian block light curve (full energy range) using the search box (e.g. by searching for the ``bblc'' string within the names of the products) and visualize it by clicking on the Preview button. Answer the question: is a flare visible?

\item Go back to the list of bulk products; select and visualize the uniform bin light curve (500 s bin, full energy range). Is a flare visible? Identify the peak and read its time, count rate and the corresponding uncertainty by clicking on the selected data point in the plot.

\item Go back to the list of bulk products; select and visualize the background light curve. Does it show any correlation between behaviour of the source and that of the background? Is there any background flare temporally coincident with the source candidate flare?

\item Go back to the list of bulk products; select and visualize the uniform bin light curve (500 s) in the super-soft, soft and hard energy ranges. Is the flare visible in all of the energy ranges? 

\item Go back to Results page, select for same observation the list of bulk products corresponding to the MOS1 camera.

\item Repeat steps 4-5 for the MOS1 camera. Is the flare visible in the Bayesian blocks light curve? In the uniform bin light curve? Is it the same feature seen with the pn camera? (simultaneous, with a similar shape of the light curve and roughly half the count rate measured by the pn?)

\item Go back to Results page, select for same observation the list of bulk products corresponding to the MOS2 camera.

\item Repeat steps 4-5 for the MOS2 camera. Is the flare visible in the Bayesian blocks light curve? In the uniform bin light curve? Is it the same feature seen with the pn camera? (same checks as for MOS1)

\end{enumerate}

The above steps 4-11 allow one to answer the major question ``Is the peculiar temporal behaviour related to some instrumental effect?''

\begin{enumerate}
\setcounter{enumi}{12}

\item From the bulk products page of any camera, go again to the 3XMM summary page and follow the link to SIMBAD results.

\item A source list is displayed, ranked by (increasing) distance from the position of the 3XMM source. Select entries within 10 arcsec from the 3XMM position by clicking on their identifier.

\item For each selected source, read the classification and take note of all the relevant information (e.g. distance, multi-wavelength counterparts, association to a Globular cluster or to a Supernova remnant...). Follow the References link, identify any published paper that could include analysis of the XMM dataset, scan it to assess if the flare is detected/described; also take note of any X-ray observation performed with other instruments. 

\item Start the ESASky app in the browser (not yet integrated with the EXTraS archive) and search the source by coordinates. A portion of the sky surrounding the selected position is visualized. It is possible to display the sky as seen at different wavelengths by different instruments.

\item Assess the presence of any counterpart at different wavelengths (near infrared, optical, gamma-ray).

\end{enumerate} 

The steps 12-16 allow one to answer the major question ``Has the source a possible counterpart at other wavelengths (e.g.  optical)?''. 

\begin{enumerate}
\setcounter{enumi}{17}

\item Using all the collected information, try to answer the last two major questions: ``What kind of astrophysical object or phenomenon might it be?'' and ``Might it be an important scientific discovery?''

\end{enumerate}

\end{appendix}

\section*{References}

\bibliographystyle{elsarticle-num}

\end{document}